\pdfoutput=1
\documentclass[sigconf,nonacm,pbalance]{acmart}
\AtEndPreamble{\hypersetup{colorlinks,linkcolor=black,citecolor=black,filecolor=black,urlcolor=ACMDarkBlue}}

\usepackage{tikz}
\definecolor{anechoic}{HTML}{08407E} %
\usepackage{enumitem}
\usepackage{placeins}
\usetikzlibrary{arrows.meta,positioning,calc,decorations.pathreplacing}
\usepackage{algorithm}
\usepackage[noend]{algpseudocode}
\algrenewcommand\algorithmicrequire{\textbf{Input:}}
\graphicspath{{images/}}
\newcommand{\fClk}{250\,MHz}
\newcommand{\ttsP}{0.050\,ms}          %
\newcommand{\ttsPr}{0.05\,ms}          %
\newcommand{\ttsS}{0.018\,ms}          %
\newcommand{\spdBSB}{5.2$\times$}      %
\newcommand{\pBoard}{108\,W}           %
\newcommand{\ets}{5.4\,mJ}             %
\newcommand{\cycStep}{18.1}            %
\newcommand{\cycFlip}{0.14}            %
\newcommand{\gpuSpd}{2.3$\times$}      %
\newcommand{\gpuEratio}{8.0$\times$}   %

\begin{document}

\title[Anechoic: Onsager-Corrected Parallel Annealing on FPGA]{Anechoic: Onsager-Corrected Parallel Annealing for Sub-0.1\,ms Dense Max-Cut on an FPGA}
\author{Seungki Hong}
\affiliation{\institution{ETH Zurich}\country{}}
\author{Kyeongwon Jeong}
\affiliation{\institution{Yonsei University}\country{}}
\author{Taekwang Jang}
\affiliation{\institution{ETH Zurich}\country{}}
\newcommand{\coderepo}{\url{https://github.com/skethz/anechoic}}

\begin{abstract}
Ising machines solve combinatorial optimization problems by annealing a network of spins. On dense problems, where every spin couples to every other, the least expensive hardware step updates all spins at once: stochastic cellular automata (SCA) decide every spin in parallel and read couplings only for the spins that flip. Such synchronous updates, however, \textit{create an echo: the previous state of each spin returns to it through its neighbors}. On K2000, the Max-Cut benchmark on a complete graph of 2{,}000 nodes, two-thirds of all flips undo a flip that the same spin made one step earlier. This paper presents \emph{Anechoic}, an FPGA Ising machine that \textit{subtracts this echo from every decision}. Dynamical mean-field theory identifies the echo as an Onsager reaction and gives the strength of its one-step part: \textit{the number of spins whose flip probability lies strictly between 0 and 1, divided by twice the temperature}. The decision logic already flags these spins, so this Onsager correction costs one population count per step, and a temperature-scaled approximation needs no count. A hand-written RTL engine executes a dense 2{,}000-spin step in \cycStep{} cycles plus \cycFlip{} cycles per flip, so every step and flip that the correction saves is saved time. In experiments, the counted and the temperature-scaled correction both find better K2000 cuts at every step budget than each of five published binary-spin algorithms run at its paper's settings. Twelve engines at \fClk{} on an AMD Alveo V80 reach a cut of at least 33{,}000 on K2000 with 99\% probability in \textbf{\ttsPr{}, \spdBSB{} faster than the fastest published hardware}, at \ets{} per solution; with 2-bit couplings, the temperature-scaled form is also faster than uncorrected SCA on 48 of 51 sparse G-set graphs. On an NVIDIA GH200 GPU, the identical engine is \gpuSpd{} slower and needs \gpuEratio{} more energy per solution, though its 240 concurrent anneals give 2.3$\times$ higher throughput. As a GlobalFoundries 22\,nm ASIC routed at 1.25\,GHz (typical corner), twelve engines would reach the target in about \textbf{10\,\textmu s, 26$\times$ faster than the fastest published hardware}, with 0.20\,mJ of engine energy per solution. The code is available at \coderepo{}.
\end{abstract}

\maketitle
\renewcommand{\shortauthors}{Hong et al.}

\section{Introduction}
\begin{figure*}[t]
  \centering
  \includegraphics[width=\textwidth]{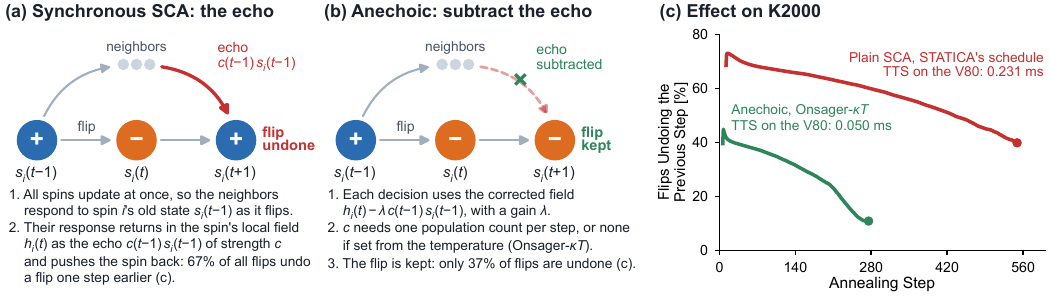}
  \caption{The echo and its removal. (a)~In synchronous SCA, the echo of a spin's state before a flip pushes the spin back. (b)~Anechoic subtracts the echo from every decision. (c)~Share of flips that undo the previous step on K2000 (engine model, 1{,}024 runs per schedule), with TTS measured on the V80 (Table~\ref{tab:configs}).}
  \Description{Three panels. Panel a shows a spin that flips between steps t minus one and t; its neighbors respond to its state before the flip, and their response returns in its local field and flips the spin back. Panel b shows the same sequence with the returning term subtracted, so the flip is kept. Panel c plots the percentage of flips that undo the previous step against the annealing step: 67 percent on average for plain SCA with STATICA's 560-step schedule, whose time-to-solution on the V80 is 0.231 ms, and 37 percent for Anechoic with 280 steps, whose time-to-solution is 0.050 ms.}
  \label{fig:headline}
\end{figure*}

Ising machines solve combinatorial optimization problems by letting a spin system relax toward low energy. Optical, analog, quantum and digital designs have all been demonstrated~\cite{doi:10.1126/science.aah4243,goto2019combinatorial,9162869,boothby2020next,9062965,10609617}. Dense problems are particularly demanding for hardware: every spin interacts with every other spin, so the problem cannot be embedded into local connectivity, and every update affects every local field. The standard benchmark for such problems is K2000, Max-Cut on a complete graph with 2{,}000 spins~\cite{doi:10.1126/science.aah4243}. Machines are compared by their time-to-solution (TTS), the time required to reach a target cut with 99\% probability. The fastest published hardware result is 0.26\,ms, from Toshiba's ballistic simulated-bifurcation machine (bSB) on an FPGA~\cite{goto2021high}.

Three families of dynamics trade work per step against progress per step: (1)~\emph{simulated annealing} (SA)~\cite{kirkpatrick1983optimization}, which updates one spin at a time and descends reliably, although on a dense graph each update rewrites the local fields of all $N$ spins, so a sweep is serial; (2)~\emph{simulated bifurcation}~\cite{goto2019combinatorial,goto2021high}, whose ballistic form (bSB) needs the fewest steps, although each step is a real-valued matrix--vector product, dense until the amplitudes settle; and (3)~\emph{synchronous stochastic cellular automata} (SCA)~\cite{9062965}, which decide all spins in parallel and access the couplings only for the spins that flip, giving the least expensive step in hardware. The limitation of SCA lies in its dynamics: a spin that flips is pushed back by the echo of its own previous state through the network, and on K2000 two-thirds of all flips reverse the flip of the previous step (Figure~\ref{fig:headline}).

This paper presents Anechoic, an FPGA Ising machine that combines this inexpensive step with a correction that subtracts the echo, as an anechoic chamber absorbs reflections (Figure~\ref{fig:headline}b). Dynamical mean-field theory~\cite{sompolinsky1982relaxational,eissfeller1992new,coolen2001statistical} identifies the echo as an Onsager reaction term, whose lag-one part has the coefficient $n_{\mathrm{lin}}/2T$, where $n_{\mathrm{lin}}$ is the number of spins whose flip probability lies strictly between 0 and 1 and $T$ is the temperature. The decision logic already flags these spins, so subtracting this lag-one term costs one population count per step. As an SCA step's cost grows with its flips, the correction saves time twice: the anneal needs fewer steps and starts colder, with fewer flips.

The contributions of this work are:
\begin{itemize}[leftmargin=*]
\item \textbf{Onsager-corrected SCA} (Section~\ref{sec:algorithm}). The echo is subtracted with a counted coefficient (Onsager-online) or a temperature-scaled one (Onsager-$\kappa T$); at every step budget on K2000, both forms reach lower Ising energy than five published binary-spin algorithms run at their papers' settings.
\item \textbf{A flip-proportional dense engine} (Section~\ref{sec:arch}). A hand-written RTL engine for the AMD Alveo V80~\cite{amd_v80} executes a dense 2{,}000-spin step in \cycStep{} cycles plus \cycFlip{} cycles per flip, bit-exact with a reference model in simulation and on the board.
\item \textbf{A measured 12-engine system} (Section~\ref{sec:evaluation}). Twelve engines reach a cut of at least 33{,}000 on K2000 with 99\% probability in \ttsP{}, \spdBSB{} faster than the fastest published hardware~\cite{goto2021high}, at \ets{} per solution. With 2-bit couplings ($-1$, 0, $+1$), Onsager-$\kappa T$ is also faster than plain SCA on 48 of 51 G-set instances.
\item \textbf{Cross-platform evidence} (Sections~\ref{subsec:gpu} and~\ref{subsec:asic}). On an NVIDIA GH200 GPU~\cite{nvidia_gh200}, the bit-exact engine is \gpuSpd{} slower and needs \gpuEratio{} more energy per solution, though its 240 concurrent anneals give higher throughput. As a 22\,nm ASIC routed at 1.25\,GHz, twelve engines would reach the target in about 10\,\textmu s, $26\times$ faster than the fastest published hardware.
\end{itemize}

\section{Related Work}\label{sec:related}
NP-hard problems such as Max-Cut, graph partitioning (GP) and the traveling salesman problem (TSP) map onto Ising machines through the choice of couplings and fields~\cite{lucas2014ising}. Anechoic is compared with the machines below, both by measured or published time-to-solution (Table~\ref{tab:comp}) and by their algorithms, which are re-implemented at their papers' settings (Figure~\ref{fig:alg}).

\noindent\textbf{Digital parallel annealers.} STATICA~\cite{9062965,yamamoto2021statica} introduces SCA, the annealing version of the probabilistic cellular automata (PCA) sampler of Dai Pra et al.~\cite{daipra2012sampling}: every spin is updated in parallel, and a self-coupling suppresses simultaneous flips. SCA with autonomous pinning-effect control (APC-SCA)~\cite{okonogi2023apc} adapts the pinning of each spin to its recent flips. Time-dimensional exchange coupling (TEC)~\cite{du2026tec} adds a coupling between consecutive time steps whose constant is grid-searched per temperature. The proposed correction has their functional form but a derived coefficient that follows the anneal (Section~\ref{subsec:onsager}). Differential stochastic simulated annealing (DSSA)~\cite{onizawa2026dssa} and graph-colored machines built from probabilistic bits (p-bits)~\cite{10185207} avoid synchronous conflicts by other means.

\noindent\textbf{Simulated bifurcation.} The simulated-bifurcation (SB) machines of Toshiba integrate classical bifurcation dynamics on FPGAs. bSB and the discrete variant (dSB)~\cite{goto2021high} extend the adiabatic SB~\cite{goto2019combinatorial}. bSB is the fastest measured hardware at the 33{,}000-cut level, and the generalized bSB (GbSB)~\cite{goto2026edge} reaches the best-known K2000~\cite{doi:10.1126/science.aah4243} cut in 9.61\,ms. Anechoic makes the opposite trade-off to their real-valued matrix--vector products, dense until the amplitudes settle~\cite{goto2021high}: a step whose cost scales with the number of flips rather than with $N^2$, at the cost of more steps or more independent trials.

\noindent\textbf{Physical, analog and software solvers.} Coherent Ising machines (CIM)~\cite{doi:10.1126/science.aah4243} and compute-in-memory designs such as ReAIM~\cite{10609617}, an adaptive Ising machine on resistive random-access memory (ReRAM), obtain the coupling product from physics or analog arrays. SA (Neal~\cite{dwave_neal}) and GPU annealers~\cite{9643932} are common software baselines.

\noindent\textbf{Onsager reaction and dynamical mean-field theory.} The Onsager reaction term originates in the Thouless--Anderson--Palmer (TAP) equations~\cite{thouless1977solution}, and approximate message passing adds an analogous lag term to deterministic iterations~\cite{bolthausen2014iterative}. Dynamical mean-field theory of disordered spin dynamics~\cite{sompolinsky1982relaxational,eissfeller1992new}, including parallel updates~\cite{coolen2001statistical}, gives the retarded (delayed) self-interaction that is identified here as the echo of synchronous annealing. Kabashima's exact synchronous Monte Carlo method~\cite{kabashima2026synchronous} finds such a retarded Onsager reaction in its analysis, but does not use it inside an annealer. To the authors' knowledge, no earlier work subtracts this term, with an online coefficient, inside a stochastic annealer.

\section{Background}\label{sec:background}

\subsection{Ising machines and Max-Cut}
The Ising model assigns spins $s_i\in\{-1,+1\}$, $i=1,\dots,N$, to the vertices of a graph with symmetric couplings $J_{ij}=J_{ji}$. For Max-Cut no external field is needed, and the Hamiltonian of a configuration $\mathbf{s}=(s_1,\dots,s_N)$ is
\begin{equation}
H(\mathbf{s}) = -\sum_{i<j} J_{ij}\, s_i s_j .
\label{eq:ising}
\end{equation}
With edge weights $w_{ij}$ and couplings $J_{ij}=-w_{ij}$, the cut of a configuration, $\mathrm{cut}(\mathbf{s})=\sum_{i<j}w_{ij}(1-s_is_j)/2$, equals $(W-H(\mathbf{s}))/2$ with $W=\sum_{i<j}w_{ij}$, so that maximizing the cut minimizes $H$. Combinatorial optimization with the Ising model, Ising machines and their annealing are introduced in~\cite{hong2026snowball}. K2000~\cite{doi:10.1126/science.aah4243} is Max-Cut on the complete graph with $N=2{,}000$ vertices and edge weights drawn uniformly from $\{-1,+1\}$; its best-known cut is 33{,}337.

\subsection{Sequential and parallel Markov chain Monte Carlo}\label{subsec:mcmc}
Simulated annealing (SA)~\cite{kirkpatrick1983optimization} runs a Markov chain while lowering its temperature $T$. Let $h_i=\sum_{j\neq i}J_{ij}s_j$ be the local field of spin $i$; flipping that spin changes the energy by $\Delta E_i=2s_ih_i$. Under Glauber dynamics~\cite{glauber1963time} the spin flips with probability $1/(1+e^{\Delta E_i/T})$. Updating one spin at a time preserves the Gibbs distribution through detailed balance~\cite{robert1999monte}.

Updating all spins simultaneously from the current fields is fully parallel, but the resulting transition no longer satisfies detailed balance. Interacting spins then flip together, and the chain oscillates between complementary configurations~\cite{heermann1990parallelization}. Synchronous Ising hardware uses three remedies: (1)~restricting parallel updates to non-interacting spins, as in graph-colored p-bit machines~\cite{10185207}, which is impossible on a complete graph; (2)~adding a self-coupling, as in PCA~\cite{daipra2012sampling} and SCA~\cite{9062965} (Section~\ref{subsec:sca}); and (3)~coupling consecutive time steps, as in TEC~\cite{du2026tec} and APC-SCA~\cite{okonogi2023apc} (Section~\ref{subsec:onsager}).

\subsection{Time-to-solution}\label{subsec:tts}
Consider a solver that reaches a target within a run of duration $t_a$ with probability $P_a$. The time required to reach the target with 99\% probability is
\begin{equation}
\mathrm{TTS}_{99} = t_a\,\frac{\ln(1-0.99)}{\ln(1-P_a)} ,
\label{eq:tts}
\end{equation}
and $\mathrm{TTS}_{99}=t_a$ when $P_a=1$; for $0.99<P_a<1$ it falls slightly below $t_a$, so a single failed run lowers it (O2 in Table~\ref{tab:configs}). Following STATICA~\cite{yamamoto2021statica} and ReAIM~\cite{10609617}, the target is a cut of at least 33{,}000 on K2000. A machine with $E$ independent engines runs in rounds: each round starts one trial, that is, one complete anneal from a random initial state, on every engine, and lasts until the slowest engine finishes. Equation~\eqref{eq:tts} can be evaluated in two ways for such a machine, and both are reported. The \emph{primary} estimator treats one round as one run. Its $t_a$ is the measured mean round time and its $P_a$ is the measured fraction of rounds in which at least one engine reaches the target. It describes the machine exactly as it operates and is the headline metric of the pre-registered protocol (Section~\ref{subsec:setup}); bSB and GbSB count batched trials the same way~\cite{goto2021high,goto2026edge}. The \emph{secondary} estimator treats every trial as a separate run. With the same $t_a$ and the measured per-trial success probability $p$, it uses $P_a=1-(1-p)^E$, which equals Equation~\eqref{eq:tts} evaluated per trial ($P_a=p$) and divided by $E$. It assumes that trials on different engines are independent, which a test on the measured trials supports, and it ignores that work is done in whole rounds; it therefore measures throughput and can fall below one round time. It is reported because, once every round succeeds, the primary estimator no longer reflects per-trial success; the two coincide for $E=1$, as for each machine in Table~\ref{tab:comp} that states how it counts runs.

\section{Onsager-Corrected Parallel Annealing}\label{sec:algorithm}

\subsection{Synchronous stochastic cellular automata}\label{subsec:sca}
The SCA of STATICA~\cite{9062965,yamamoto2021statica} anneals the PCA sampler of Dai Pra et al.~\cite{daipra2012sampling}, which updates all spins simultaneously and whose stationary distribution approaches the Gibbs distribution as the self-coupling and the number of spins grow. With $h_i(t)$ the local field of spin $i$ at step $t$, spin $i$ keeps its value with probability
\begin{equation}
  P\bigl(s_i(t{+}1)=s_i(t)\bigr)=\operatorname{clamp}\!\left(\frac{z_i(t)}{4T_t}+\frac12,\,0,\,1\right),
  \label{eq:sca}
\end{equation}
where $z_i(t)=s_i(t)\,h_i(t)+q$, $\operatorname{clamp}(x,a,b)=\max(a,\min(x,b))$, $T_t$ is the temperature at step $t$, and $q\geq0$ is a self-coupling that pins spins and suppresses simultaneous flips. An anneal of $S$ steps lowers $T_t$ geometrically from an initial temperature $T_0$ to a final temperature of 5. A decision is called \emph{linear} when $|z_i(t)|<2T_t$, that is, when its probability lies strictly between 0 and 1; outside this band the decision is deterministic. Each decision requires one field, one comparison with a uniform random threshold and no \mbox{exponential}, unlike the Glauber dynamics of SA (Section~\ref{subsec:mcmc}), and all $N$ decisions are independent given the current state, so a step parallelizes completely. In exchange, synchronous updates violate detailed balance (Section~\ref{subsec:mcmc}), and part of the dynamics oscillates instead of descending, as flips are undone one step later (Figure~\ref{fig:headline}a).

\subsection{The echo of synchronous updates}\label{subsec:echo}
The oscillation comes from the couplings: each neighbor $j$ chooses its state $s_j(t)$ from a field that contains $J_{ji}s_i(t{-}1)$. The field $h_i(t)=\sum_jJ_{ij}s_j(t)$ therefore contains a component along $s_i(t{-}1)$: the network returns the previous state of spin $i$ to that spin, and if the spin has just flipped, this component pushes it back (Figure~\ref{fig:headline}a). On K2000~\cite{doi:10.1126/science.aah4243} with STATICA's short schedule~\cite{yamamoto2021statica} ($q=4$, $T_0=30$, $S=560$; B5 in Table~\ref{tab:configs}), 67\% of all flips in 1{,}024 simulated runs reverse a flip that the same spin made one step earlier (Figure~\ref{fig:headline}c, upper curve).

\begin{figure}[t]
  \centering
  \includegraphics[width=\columnwidth]{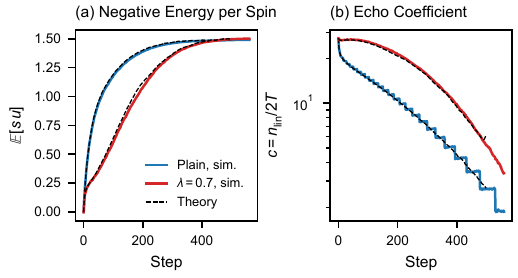}
  \caption{Dynamical mean-field theory~\cite{sompolinsky1982relaxational,eissfeller1992new,coolen2001statistical} (dashed) versus simulation of K2000 (256 runs; $T_0=30$, $S=560$, $q=4$) for plain SCA~\cite{9062965} and Onsager-online ($\lambda=0.7$): (a)~negative energy per spin $\mathbb{E}[s\,u]=-2H/N^{3/2}$, with $u=h/\sqrt{N}$; (b)~echo coefficient $c=n_{\mathrm{lin}}/2T$. Up to step 500, where the theory stops because its solver loses accuracy, theory and simulation differ by at most 0.06 in $\mathbb{E}[s\,u]$ and by 3\% in $c$ on average.}
  \Description{Two panels. Left: the negative energy per spin rises from 0 to about 1.5 over 560 steps; simulation and theory overlap for both rules. Right: the echo coefficient falls by about an order of magnitude over the anneal, with theory and simulation overlapping.}
  \label{fig:theory}
\vspace{-0.6em}%
\end{figure}

The echo is quantified with dynamical mean-field theory, the standard large-$N$ analysis of disordered spin dynamics~\cite{sompolinsky1982relaxational,eissfeller1992new,coolen2001statistical}. Let $u=h/\sqrt{N}$ and $\tilde T=T/\sqrt{N}$ denote the field and the temperature rescaled by $\sqrt{N}$, the typical magnitude of a sum of $N$ random $\pm1$ couplings. For $N\to\infty$, averaging over symmetric random couplings reduces the parallel dynamics to a single effective spin $s(t)$ whose field is
\begin{equation}
  u(t)=\eta(t)+\sum_{t'<t}G(t,t')\,s(t'),
  \label{eq:dmft}
\end{equation}
where the Gaussian process $\eta$ has covariance $C(t,t')=\mathbb{E}[s(t)s(t')]$ and $G(t,t')=\partial\mathbb{E}[s(t)]/\partial\theta(t')$ is the response of the mean spin to a small field perturbation $\theta$ applied at step $t'$. The second term of Equation~\eqref{eq:dmft} is the retarded Onsager reaction~\cite{thouless1977solution}, referred to here as the echo. Its lag-one part has a direct reading. Spin $i$ enters the field of each neighbor $j$ through $J_{ji}$; a neighbor whose decision is linear changes its expected next state by $1/(2T_t)$ per unit of field (F1, F2 in Table~\ref{tab:formal}), and any other neighbor not at all; and the change returns to spin $i$ through $J_{ij}$, with $J_{ij}J_{ji}=1$ on K2000. The lag-one echo in $h_i(t{+}1)$ is therefore $c(t)\,s_i(t)$, with
\begin{equation}
  c(t)=\frac{n_{\mathrm{lin}}(t)}{2T_t},
  \label{eq:c}
\end{equation}
where $n_{\mathrm{lin}}(t)$ is the number of spins whose decision at step $t$ is linear. Its run-to-run fluctuations vanish as $N\to\infty$, and at finite $N$ the online count follows each run's own fluctuations. Even at $N=2{,}000$ the theory matches direct simulation of K2000 to within 3\% in $c$ on average (Figure~\ref{fig:theory}).

The finite identities used here and in the engine are machine-checked twice, independently, in Lean~4~\cite{demoura2021lean4} with Mathlib~\cite{mathlib2020} and in the Rocq Prover~\cite{rocq2026} (Table~\ref{tab:formal}), with no unproved step and only the libraries' standard axioms; the large-$N$ step is not formalized.

\begin{table}[t]
\centering
\caption{Statements machine-checked in both Lean~4~\cite{demoura2021lean4} with Mathlib~\cite{mathlib2020} and the Rocq Prover~\cite{rocq2026}. Here $f(g)=\operatorname{clamp}(g/2T,-1,1)$ is the expected next spin at field $g=s_iz_i$.}
\label{tab:formal}
\setlength{\tabcolsep}{3pt}
\begin{tabular}{@{}lp{0.80\columnwidth}@{}}
\toprule
& Statement \\
\midrule
F1 & Eq.~\eqref{eq:sca} gives $P(s_i(t{+}1)={+}1)=\bigl(1+f(h_i+q\,s_i)\bigr)/2$ \\
F2 & $f'(g)=1/(2T)$ for $|g|<2T$ and $f'(g)=0$ for $|g|>2T$; hence $\sum_i f'(g_i)=n_{\mathrm{lin}}/(2T)$ \\
F3 & $\sqrt{N}\,\bigl(n_{\mathrm{lin}}/N\bigr)/\bigl(2T/\sqrt{N}\bigr)=n_{\mathrm{lin}}/(2T)$ (rescaled to unscaled units) \\
F4 & $s(h-\lambda c\,s')+q=s\,h+(q-\lambda c\,s\,s')$ ($q_{\mathrm{eff}}$ identity) \\
F5 & $\sum_e 2\sigma_eJ_e=4\,\#\{e: J_e=\sigma_e\}-2\,\#E$ (population-count field update) \\
F6 & $(256k+32m+4a+e)\bmod 4=e$ for $m,a<8$, $e<4$ (and modulo 8 for eight extractors); the representation is unique \\
F7 & $\mathrm{cut}(\mathbf{s})=(W-H(\mathbf{s}))/2$ \\
\bottomrule
\end{tabular}
\vspace{-0.6em}%
\end{table}

\subsection{Onsager correction and its hardware cost}\label{subsec:onsager}
\begin{algorithm}[t]
\caption{One anneal of Onsager-online ($s'$: previous state). Onsager-$\kappa T$ uses $\beta\gets\kappa\,T_{t-1}\rho_t$ in line~3.}
\label{alg:anneal}
\begin{algorithmic}[1]
\Require $J$, $q$, and $T_t$, $\lambda_t$ for $t<S$
\State $s\gets$ random spins;\; $s'\gets s$;\; $h\gets Js$;\; $n_{\mathrm{lin}}\gets0$
\For{$t\gets0$ \textbf{to} $S-1$}
  \State $\beta\gets\lambda_t\,n_{\mathrm{lin}}/(2T_{t-1})$, or $\beta\gets0$ if $t=0$
  \State $n_{\mathrm{lin}}\gets0$
  \ForAll{spins $i$ \textbf{in parallel}}
    \State $z_i\gets s_i\,(h_i-\beta\,s'_i)+q$
    \State \textbf{if} $|z_i|<2T_t$ \textbf{then} $n_{\mathrm{lin}}\gets n_{\mathrm{lin}}+1$
    \State $f_i\gets[\,z_i<r_i\,]$ with $r_i$ uniform in $[-2T_t,2T_t)$
  \EndFor
  \State $s'\gets s$;\; $s_x\gets-s_x$ for every $x$ with $f_x=1$
  \ForAll{$x$ with $f_x=1$} \Comment{only these rows of $J$}
    \State $h_y\gets h_y+2\,s_xJ_{xy}$ for all $y$
  \EndFor
\EndFor
\State \Return $s$
\end{algorithmic}
\end{algorithm}
The estimated lag-one echo is subtracted from the decision field (Figure~\ref{fig:headline}b):
\begin{equation}
  z_i(t)=s_i(t)\bigl[h_i(t)-\lambda_t\,c(t{-}1)\,s_i(t{-}1)\bigr]+q ,
  \label{eq:onsager}
\end{equation}
and the rule of Equation~\eqref{eq:sca} is otherwise unchanged. The gain $\lambda_t=\lambda\rho_t$ combines a constant $\lambda$ (0.7--1.05 in the schedules used here) with an optional end ramp $\rho_t$, which equals 1 for the first 70\% of the steps and decreases linearly to 0 over the remaining 30\%; with the ramp, the anneal ends as plain SCA~\cite{9062965} and freezes.

Rewriting Equation~\eqref{eq:onsager} as $s_ih_i+q_{\mathrm{eff},i}$, with $q_{\mathrm{eff},i}=q-\lambda_t\,c(t{-}1)\,s_i(t)s_i(t{-}1)$, reveals the mechanism: a spin that has not flipped loses pinning, and a spin that has just flipped gains it. TEC~\cite{du2026tec} and the adaptive per-spin pinning of APC-SCA~\cite{okonogi2023apc} have the same functional form; the theory contributes the coefficient, which must follow $c(t)$ over more than an order of magnitude during an anneal, with a fixed sign. A constant coupling cannot; TEC's best constant indeed drifts with temperature.

With $\lambda=0.7$, the correction reduces immediate flip-backs from 67\% to 54\% of all flips. It also lets the anneal start much colder: the same selection procedure (Section~\ref{subsec:setup}) picks $T_0=12$--$15$ for both forms but 30--40 for plain SCA, although its grid includes 15. A colder start shortens the flip-intensive early phase, which matters for the hardware, whose cost grows with the number of flips (Section~\ref{sec:arch}). The theory also predicts when the correction is stable, and three such predictions, each stated before the corresponding experiment, hold on K2000 (1{,}024 runs each, $T_0=30$, $S=560$): (1)~at $q=4$, full cancellation ($\lambda=1$) prevents the spins from ordering (mean cut 9{,}721); (2)~at $q=4$, there is a stability threshold $\lambda_c\approx0.8$--$0.9$ (success 0.61 at $\lambda=0.8$, 0.24 at 0.9); and (3)~$\lambda_c$ increases with $q$ (success 0.63 at $\lambda=0.9$ with $q=8$).

The correction is evaluated in two forms, named after how the coefficient is obtained. \emph{Onsager-online} uses the counted coefficient of Equation~\eqref{eq:c}. \emph{Onsager-$\kappa T$} replaces $\lambda\,c(t{-}1)$ by $\kappa T_{t-1}$, where $\kappa$ is a tuned constant:
\begin{equation*}
  z_i(t)=s_i(t)\bigl[h_i(t)-\kappa\,T_{t-1}\,\rho_t\,s_i(t{-}1)\bigr]+q .
\end{equation*}
For plain SCA on K2000, $n_{\mathrm{lin}}/4T^2$ stays near 0.3 for $8\le T\le25$, as the theory predicts, so $c=n_{\mathrm{lin}}/2T$ is nearly proportional to $T$ (Figure~\ref{fig:theory}b), and a coupling $\kappa T$ follows it without a population count. Onsager-$\kappa T$ retains the gain: in pre-registered ablations at equal hardware cost, the online coefficient gives a further 4--9\% speed-up in projected time-to-solution, but on the board the best Onsager-$\kappa T$ schedule is the faster one (Section~\ref{subsec:ttsres}).

The correction is inexpensive: $n_{\mathrm{lin}}$ is a population count of flags that the lanes already compute, $\lambda_t/(2T_{t-1})$ is a table entry, and each lane adds or subtracts their product $\lambda_t c(t{-}1)$ depending on whether $s_i(t)=s_i(t{-}1)$. Both forms run on the same engine and differ only in two per-step tables (Algorithm~\ref{alg:anneal}).

\begin{figure*}[t]
  \centering
  \includegraphics[width=\textwidth]{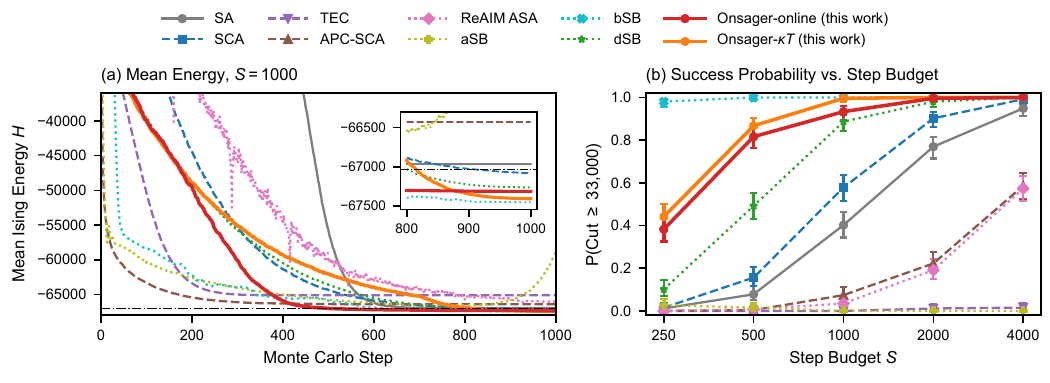}
  \caption{Algorithm-level comparison on K2000. (a)~Mean energy versus Monte Carlo step at $S=1000$ (inset: last 20\% of steps; dash-dotted: 33{,}000-cut target). (b)~Success probability versus step budget over 256 held-out runs, with 95\% Wilson intervals~\cite{wilson1927probable}. Baselines use their papers' settings; only the corrected forms are tuned. aSB~\cite{goto2019combinatorial} diverges at long budgets.}
  \Description{Two panels with a shared legend. Left: energy decreases with Monte Carlo steps for ten methods; the two corrected SCA rules and simulated bifurcation reach the lowest energies. Right: success probability rises with the step budget; the two forms of the correction dominate the other binary-spin methods.}
  \label{fig:alg}
\vspace{-0.6em}%
\end{figure*}

\subsection{Comparison with published algorithms}\label{subsec:algcomp}
Figure~\ref{fig:alg} compares the two forms of the correction by energy versus Monte Carlo step with (1)~the algorithms of the Ising machines in Table~\ref{tab:comp}, namely the SCA of STATICA~\cite{9062965}, the adaptive solver algorithm (ASA) of ReAIM~\cite{10609617} without ReRAM noise, and adiabatic simulated bifurcation (aSB)~\cite{goto2019combinatorial}, bSB and dSB~\cite{goto2021high}; (2)~two other corrections that couple consecutive time steps, TEC~\cite{du2026tec} and APC-SCA~\cite{okonogi2023apc}; and (3)~classical SA~\cite{kirkpatrick1983optimization} with sequential Metropolis sweeps~\cite{metropolis1953equation}.
One step gives every spin one update opportunity: (1)~for SA and TEC, a sweep of $N$ sequential updates; (2)~for the SCA family, one parallel update; (3)~for SB, one integration step, which requires one dense product of $J$ with the amplitude vector; and (4)~for ASA, one iteration.
Every baseline uses the settings that its paper describes: Neal's default temperatures for SA~\cite{dwave_neal}; $q=4$ and $T$ from 40 to 5 for SCA~\cite{yamamoto2021statica}; the reset, rates and temperatures of APC-SCA~\cite{okonogi2023apc}; the temperatures, $F$ and $k$ of ASA~\cite{10609617}; $J_v=30J$ and $T$ from $100J$ to $0.1J$ for TEC~\cite{du2026tec}; and the step sizes and scalings that their papers prescribe for aSB, bSB and dSB~\cite{goto2019combinatorial,goto2021high}. Each implementation was checked against results that its paper reports. Only TEC is not reproduced: implemented with sequential p-bit updates, it reaches the paper's 95\% target in 577 instead of 740 sweeps but gains only 1.28 times, not 1.88 times, from its coupling. Only the two forms of the correction are tuned: for each step budget $S\in\{250,500,1000,2000,4000\}$, the configuration with the highest mean final cut over 64 tuning runs is selected and then evaluated on 256 held-out runs with fresh seeds.

Among the binary-spin methods, that is, all except the three SB variants, whose amplitudes are continuous, the two forms of the correction reach the lowest mean energy at every budget, including sequential SA. At $S=500$, they reach the 33{,}000 target in 87\% (Onsager-$\kappa T$) and 82\% (Onsager-online) of the runs, against 16\% for SCA~\cite{9062965}, 8\% for SA~\cite{kirkpatrick1983optimization}, and at most 1\% for TEC~\cite{du2026tec}, APC-SCA~\cite{okonogi2023apc} and ASA~\cite{10609617}. A second measure is the number of steps to solution, $S\cdot\ln0.01/\ln(1-p)$ minimized over $S$, where $p$ is the fraction of runs that reach the target. By this measure, Onsager-$\kappa T$ needs 830 steps and Onsager-online 1{,}358. dSB needs 2{,}115, SCA 3{,}796, SA 6{,}181, APC-SCA and ASA about 21{,}000, aSB 41{,}526 and TEC about 780{,}000. Only bSB~\cite{goto2021high} needs fewer (293).

bSB also reaches the lowest energy for $S\leq2000$. It then stalls, and at $S=4000$ both forms end at lower energy (mean cuts 33{,}236 and 33{,}233; bSB 33{,}209, dSB 33{,}215). Section~\ref{subsec:sota} explains why SCA's cheaper steps give a lower measured TTS than bSB.

\section{Engine and System Architecture}\label{sec:arch}

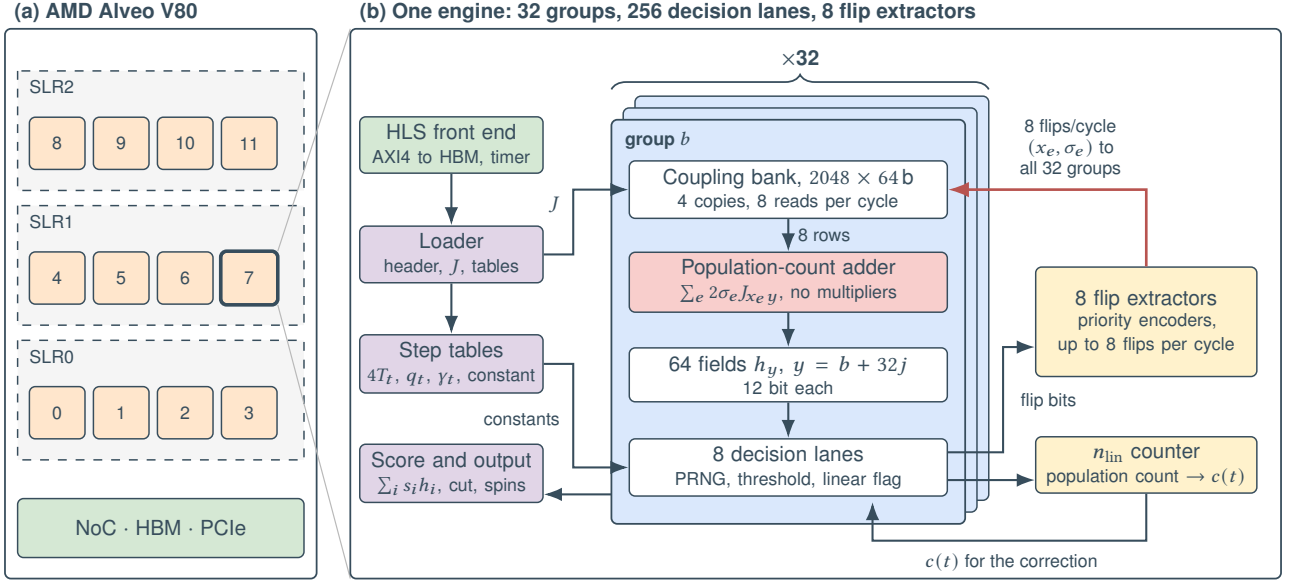
\begin{figure*}[t]
\centering
\definecolor{figtext}{HTML}{374D5C}\definecolor{figblue}{HTML}{DAE8FC}\definecolor{figgreen}{HTML}{D5E8D4}%
\definecolor{figorange}{HTML}{FFE6CC}\definecolor{figpurple}{HTML}{E1D5E7}\definecolor{figred}{HTML}{F8CECC}\definecolor{figredline}{HTML}{B85450}\definecolor{figyellow}{HTML}{FFF2CC}\definecolor{figgray}{HTML}{F5F5F5}%
\resizebox{0.95\textwidth}{!}{%
\begin{tikzpicture}[
  font=\fontfamily{phv}\selectfont\footnotesize, >=Latex, text=figtext,
  blk/.style={draw=figtext, rounded corners=2pt, align=center, inner sep=2.5pt, fill=white, line width=0.5pt},
  lab/.style={align=center, font=\fontfamily{phv}\selectfont\scriptsize},
  bus/.style={->, line width=0.7pt, draw=figtext},
  bc/.style={->, line width=0.9pt, draw=figredline}
]
\node[anchor=north west, font=\fontfamily{phv}\selectfont\footnotesize\bfseries] at (0,0.3) {(a) AMD Alveo V80};
\draw[figtext, line width=0.6pt, rounded corners=2pt, fill=white] (0,-0.12) rectangle (3.7,-6.65);
\foreach \s/\k in {2/0,1/1,0/2} {
  \pgfmathsetmacro{\yt}{-0.62-\k*1.6}
  \draw[figtext, dashed, line width=0.5pt, fill=figgray] (0.15,\yt) rectangle (3.55,\yt-1.42);
  \node[anchor=north west, lab] at (0.18,\yt) {SLR\s};
  \foreach \e in {0,1,2,3} {
    \pgfmathtruncatemacro{\n}{4*\s+\e}
    \node[blk, fill=figorange, minimum width=0.66cm, minimum height=0.62cm, lab] (E\n) at (0.62+\e*0.76,\yt-0.86) {\n};
  }
}
\node[blk, fill=figgreen, minimum width=3.4cm, minimum height=0.7cm] at (1.85,-6.05) {NoC $\cdot$ HBM $\cdot$ PCIe};
\draw[figtext, line width=1.2pt, rounded corners=2pt] (E7.south west) rectangle (E7.north east);
\draw[gray!55, line width=0.4pt] (E7.north east) -- (4.1,-0.12);
\draw[gray!55, line width=0.4pt] (E7.south east) -- (4.1,-6.65);
\node[anchor=north west, font=\fontfamily{phv}\selectfont\footnotesize\bfseries] at (4.1,0.3) {(b) One engine: 32 groups, 256 decision lanes, 8 flip extractors};
\draw[figtext, line width=0.6pt, rounded corners=2pt, fill=white] (4.1,-0.12) rectangle (15.15,-6.65);
\node[blk, fill=figgreen, text width=2.0cm] (fe) at (5.3,-1.5) {HLS front end\\{\scriptsize AXI4 to HBM, timer}};
\node[blk, fill=figpurple, text width=2.0cm] (ld) at (5.3,-2.8) {Loader\\{\scriptsize header, $J$, tables}};
\node[blk, fill=figpurple, text width=2.0cm] (tab) at (5.3,-4.1) {Step tables\\{\scriptsize $4T_t$, $q_t$, $\gamma_t$, constant}};
\node[blk, fill=figpurple, text width=2.0cm] (sc) at (5.3,-5.4) {Score and output\\{\scriptsize $\sum_i s_ih_i$, cut, spins}};
\draw[bus] (fe) -- (ld);
\draw[bus] (ld) -- (tab);
\foreach \o in {0.28,0.14} \draw[figtext, line width=0.5pt, rounded corners=2pt, fill=figblue] (7.2+\o,-1.2+\o) rectangle (11.4+\o,-6.0+\o);
\draw[figtext, line width=0.6pt, rounded corners=2pt, fill=figblue] (7.2,-1.2) rectangle (11.4,-6.0);
\node[anchor=north west, font=\fontfamily{phv}\selectfont\scriptsize\bfseries] at (7.25,-1.23) {group $b$};
\draw[figtext, line width=0.6pt, decorate, decoration={brace, amplitude=5pt}] (7.2,-0.84) -- (11.68,-0.84);
\node[anchor=south, font=\fontfamily{phv}\selectfont\footnotesize\bfseries] at (9.44,-0.66) {$\times$32};
\node[blk, text width=3.6cm] (cb) at (9.3,-2.03) {Coupling bank, $2048\times64$\,b\\{\scriptsize 4 copies, 8 reads per cycle}};
\node[blk, fill=figred, text width=3.6cm] (pc) at (9.3,-3.13) {Population-count adder\\{\scriptsize $\sum_e 2\sigma_e J_{x_e y}$, no multipliers}};
\node[blk, text width=3.6cm] (fd) at (9.3,-4.23) {64 fields $h_y$, $y=b+32j$\\{\scriptsize 12 bit each}};
\node[blk, text width=3.6cm] (ln) at (9.3,-5.33) {8 decision lanes\\{\scriptsize PRNG, threshold, linear flag}};
\draw[bus] (cb) -- node[right, lab]{8 rows} (pc);
\draw[bus] (pc) -- (fd);
\draw[bus] (fd) -- (ln);
\draw[bus] (ld.east) -- ++(0.35,0) |- (cb.west);
\node[lab, anchor=east] at (6.7,-2.2) {$J$};
\draw[bus] (tab.east) -- ++(0.35,0) |- (ln.west);
\node[lab, anchor=east] at (6.7,-4.75) {constants};
\draw[bus] (7.2,-5.65) -- (sc.east|-0,-5.65);
\node[blk, fill=figyellow, text width=2.45cm, minimum height=1.3cm] (ex) at (13.55,-3.6) {8 flip extractors\\{\scriptsize priority encoders,\\up to 8 flips per cycle}};
\node[blk, fill=figyellow, text width=2.45cm] (nl) at (13.55,-5.3) {$n_{\mathrm{lin}}$ counter\\{\scriptsize population count $\to c(t)$}};
\draw[bus] (11.19,-5.15) -- (11.85,-5.15) -- (11.85,-3.9) -- (ex.west|-0,-3.9);
\node[lab, anchor=west] at (11.95,-4.55) {flip bits};
\draw[bus] (11.19,-5.48) -- (nl.west|-0,-5.48);
\draw[bus] (nl.south) -- (13.55,-6.2) -- (10.3,-6.2) -- (10.3,-5.74);
\node[lab, anchor=north] at (11.95,-6.24) {$c(t)$ for the correction};
\draw[bc] (ex.north) -- (13.55,-2.03) -- (cb.east);
\node[lab, anchor=south] at (12.65,-2.0) {8 flips/cycle\\$(x_e,\sigma_e)$ to\\all 32 groups};
\end{tikzpicture}}
\caption{(a)~Twelve engines, four per SLR, share the NoC and HBM of the AVED shell~\cite{amd_aved}, which connects to the host through PCIe. (b)~One engine. Each of the 32 groups (one shown) owns a coupling bank, 64 fields $h_y$ with $y=b+32j$ and 8 decision lanes, so coupling reads and field updates stay in the group. Eight extractors drain the flip bits while later rounds are still deciding and broadcast up to eight flipped spins $(x_e,\sigma_e)$ per cycle to all groups, which add $2\sigma_e J_{x_e y}$ to their fields by population counts. The count $n_{\mathrm{lin}}$ of linear decisions gives the coefficient $c(t)$ of the correction. PRNG: pseudorandom number generator.}
\Description{Left: a V80 device with three super logic regions, each holding four numbered engines, and a NoC, HBM and PCIe bar; engine 7 is highlighted and connected by zoom lines to the right panel. Right: one engine, with a front end, loader, step tables and score unit on the left; a stack of 32 group cards in the middle, the front card showing a coupling bank, a population-count adder, 64 fields and 8 decision lanes connected top to bottom; on the right, eight flip extractors that broadcast flipped spins back to the coupling banks of all groups, and an n_lin counter that returns the correction coefficient to the lanes.}
\label{fig:arch}
\vspace{-0.6em}%
\end{figure*}

Each step costs only what the algorithm needs: $N$ decisions, plus one coupling row for each spin that flips. All spin and field state lives on chip exactly once, and the datapath contains no handshakes. Earlier versions (v1--v3 in Table~\ref{tab:evo}) are written in high-level synthesis (HLS), whose duplicated loop state and handshakes limit timing at 250--300\,MHz. This engine is hand-written RTL, and its only HLS part is a small front end that moves data (Figure~\ref{fig:arch}).

\subsection{Data layout and decisions}\label{subsec:layout}
$J_{ij}\in\{\pm1\}$ is stored as one bit per entry. The padded $2048\times2048$ matrix is split by columns into 32 banks. Word $x$ of bank $b$ holds the 64 couplings $J_{x,\,b+32j}$ ($j=0..63$). Each bank exists four times, two copies in block random-access memory (BRAM) and two in UltraRAM (URAM), and each copy has two ports, which gives eight row reads per bank per cycle (16\,Mbit per engine). Group $b$ owns the 64 fields $h_y$ ($y=b+32j$, 12-bit), their spins, and eight decision lanes. Lane $l$ decides spin $l+256k$ in round $k$, so a step takes eight rounds of 256 parallel decisions.

Each lane has its own xoshiro128** generator~\cite{blackman2021scrambled}. At the start of a trial the generators are seeded through a 256-stage shift chain from one pipelined splitmix64 unit~\cite{steele2014splitmix}, which avoids a seed broadcast. A lane evaluates $z$ of Equation~\eqref{eq:onsager} in Q16.16 fixed point (16 integer and 16 fractional bits) and compares it with the threshold $((r-2^{15})\cdot4T_t)\gg16$, where $r$ is a 16-bit random number and $\gg$ denotes an arithmetic right shift. This needs one digital signal processing (DSP) multiplier per lane. The lane also reports whether the decision is linear ($|z|<2T_t$). Both comparisons are split into exact 16-bit halves to shorten the carry chains. The per-step correction is $(n_{\mathrm{lin}}(t{-}1)\cdot\gamma_t)\gg8$, where $\gamma_t=\lambda_t/(2T_{t-1})$ is stored in Q8.24 format (8 integer and 24 fractional bits). A second table entry holds the constant term of Onsager-$\kappa T$ or of TEC's coupling added to SCA~\cite{du2026tec}.

\subsection{Flip extraction and field update}\label{subsec:update}
Eight extractors each drain one 32-bit word of flip bits per round, eight lanes from each of four groups, so together they cover the 256 lanes of a round. A priority encoder emits one flipped spin $x_e$ and its new sign $\sigma_e$ per cycle per extractor. Extraction of round $k$ starts while rounds $k{+}1..7$ are still deciding. Each cycle the up to eight flips are broadcast to all groups. Every group reads the eight coupling words $J_{x_e,\cdot}$, one per bank port, and updates each of its 64 fields by $\sum_e 2\sigma_eJ_{x_e y}$. That sum is an 8-input population count followed by a 6-bit addition, so 16{,}000 coupling terms, eight for each of the 2{,}048 fields, are applied per cycle without a single multiplier. Field updates become visible six cycles after issue; the next round reads its fields two cycles after the extractors drain, the shortest bit-exact latency.

The resulting cost of one trial in clock cycles, fitted on RTL simulation of one trial for each of six protocol configurations, is
\begin{equation}
  \text{cycles}=\cycFlip\cdot\text{flips}+\cycStep\cdot S+886 ,
  \label{eq:cycles}
\end{equation}
where flips is the total number of spin flips in the trial and $S$ is the number of steps. Each step costs \cycStep{} cycles to issue its eight decision rounds and drain the pipeline. Each flip adds \cycFlip{} cycles, slightly more than the ideal $1/8$, because a step waits for the busiest of the eight extractors. The constant covers the per-trial work outside the steps: seeding the generators, drawing the random initial state and building its fields, computing $\sum_is_ih_i$ and the cut in a pipelined adder tree, and streaming the result out. For example, a trial with $S=280$ and about 47{,}400 flips takes about 12{,}700 cycles in the model and 12{,}371 on the board on average (49\,\textmu s at 250\,MHz).
On the board, the fastest schedules average 38--44 cycles per step (153--177\,ns). Eight extractors and four coupling copies instead of four and two cut the cycles per trial by 28--32\% for 15\% more lookup tables (LUTs) and identical results (Table~\ref{tab:evo}).

\subsection{Front end and multi-engine system}\label{subsec:system}
An HLS front end gives each engine two AXI4 masters into the high-bandwidth memory (HBM) through the network on chip (NoC): a 256-bit master for couplings and tables and a second master for results. Couplings and tables stay resident across launches, so after the first launch a launch moves only a two-beat header.

The system places twelve engines, four per super logic region (SLR), with one placement block per SLR. The engines connect through two SmartConnects to the NoC of the Alveo Versal Example Design (AVED) shell~\cite{amd_aved}, which also provides the processing system, the PCIe interface and HBM. The kernel clock is 250\,MHz. Timing closure requires six changes: (1)~replicated clock enables and per-step constants, (2)~the split comparisons, (3)~a registered coupling-address multiplexer, (4)~a pipelined score tree, (5)~an input skid buffer, and (6)~post-route physical optimization.
These move the stand-alone four-extractor engine (v4) from $-0.78$\,ns to $+0.32$\,ns of slack at 300\,MHz; the eight-extractor engine alone meets 275\,MHz.

\subsection{Verification}\label{subsec:verif}
The engine reproduces a C++ reference model bit for bit: spins, cut, flip count and the per-step $n_{\mathrm{lin}}$ trace. This holds in RTL simulation for every protocol configuration. On the board, each image must first pass an exact gate. Seeded trials on every engine, with per-step traces, are compared bit for bit against the reference (60/60 on each 12-engine version). Because trial outcomes depend only on the seed and the trial index, images must also agree with each other. Versions v4--v6 (Table~\ref{tab:evo}) return identical results for all 41{,}040 common trials, and v7 for all 4{,}164 trials it shares with v6.

\subsection{Multi-bit couplings and bias}\label{subsec:multibit}
Sparse or weighted graphs need couplings beyond $\pm1$, and encodings with linear terms need a per-spin bias~\cite{lucas2014ising}. The multi-bit engine (v7 in Table~\ref{tab:evo}) stores $K$-bit couplings $J_{xy}\in\{-M,\dots,M\}$, $M=2^{K-1}-1$, as a sign plane and $K{-}1$ magnitude planes in the bank layout of Section~\ref{subsec:layout}. With $\tau_e=\sigma_e\operatorname{sign}(J_{x_ey})$ and $m_{be}$ bit $b$ of $|J_{x_ey}|$, the field update $\sum_e2\sigma_eJ_{x_ey}$ of Section~\ref{subsec:update} becomes $2\sum_b2^b\bigl(2\,\#\{e:m_{be}\wedge\tau_e{=}{+}1\}-\#\{e:m_{be}\}\bigr)$: two 8-input population counts per magnitude plane and still no multiplier. Each coupling is stored once, in a single-port bank per extractor, because extractor $e$ applies only flips of spins $x$ with $x\bmod8=e$. At $K=2$ this halves the stored bits of v6 (8 instead of 16\,Mbit per engine), but the store is limited by read ports rather than capacity: the engine keeps the BRAM of v6 and doubles its URAM. A bias $b_i$ enters only the initial fields, $h_i(0)=b_i+\sum_jJ_{ij}s_j(0)$; the incremental update preserves it, so it costs no cycles per step, and a runtime problem size $n\leq2{,}048$ masks unused spins. Twelve engines with $K=2$ close timing at 250\,MHz with 89\% of the LUTs and 81\% of the URAM. With $K=4$, the memory allows six engines, which fail routing at 250\,MHz but close timing at 225\,MHz (86\% of the LUTs). The $K=2$ engine is bit-exact with the reference model in simulation and on the board, including runtime sizes of 800--2{,}000 spins, and on K2000 it returns the same trials as v6 in the same cycles. With bias, the six $K=4$ engines are bit-exact on the board in 114 of 114 trials, including GP and TSP encodings, as is the GPU kernel (Section~\ref{subsec:gpu}; 6{,}308 trials).

\section{Evaluation}\label{sec:evaluation}

\subsection{Experimental methodology}\label{subsec:setup}
\textbf{Platform.} Anechoic runs on an AMD Alveo V80 (Versal HBM, xcv80, three SLRs)~\cite{amd_v80} with the AVED shell~\cite{amd_aved}, implemented with Vivado 2025.1. Twelve eight-extractor engines run at \fClk{}.

\noindent\textbf{Timing.} Each engine's start-to-done cycle counter, divided by the kernel clock, gives the device time. Mean round times exclude the first round, which also loads the couplings (0.58\,ms). For 64-trial launches, host wall-clock time agrees with the counters to within 0.3\%.

\noindent\textbf{Problem and scoring.} Unless stated otherwise, the hardware experiments use K2000~\cite{doi:10.1126/science.aah4243} (Section~\ref{sec:background}), as do all works in Table~\ref{tab:comp}. A trial succeeds if its final state has a cut of at least 33{,}000. The host rescores every returned state independently of the device.

\noindent\textbf{Protocol.} A configuration, or schedule, specifies the rule, $q$, $\lambda$ or $\kappa$, $T_0$ and $S$. The configurations, seeds, trial counts, estimators and quantitative predictions are frozen and hashed with SHA-256 before any data are taken. Later configurations enter only through amendments frozen in the same way. Each configuration runs 2{,}052 trials, that is, 171 rounds of 12 engines. Nine Onsager schedules of both forms are measured: five are re-selected with the earlier v5 cycle model (Table~\ref{tab:evo}), four for nine engines and O1 for twelve, and checked on 2{,}048 held-out runs, and four come from an earlier ablation. The baselines are STATICA's two published K2000 schedules for plain SCA~\cite{yamamoto2021statica} (B5, B7) and, for reference, this work's plain-SCA (B2, B3, B6) and TEC-style schedules (B1, B4), which add TEC's coupling~\cite{du2026tec} to SCA.
Every pre-registered K2000 prediction for the 12-engine system holds, including time (0.051\,ms predicted, 0.050\,ms measured for O1). On the earlier 9-engine v4, the time predictions fail because the model omits per-launch table streaming, which motivated the resident tables (Section~\ref{subsec:system}).

\subsection{Time-to-solution}\label{subsec:ttsres}
\begin{table}[t]
\centering
\caption{Measured TTS$_{99}$ [ms] on the V80 (12 engines, 250\,MHz, cut $\geq$ 33{,}000, 2{,}052 trials each). $t_r$: mean round time; primary: from rounds; secondary: from the per-trial success probability $p$ (Section~\ref{subsec:tts}). O: Onsager schedules; B: baselines, with STATICA's published K2000 schedules~\cite{yamamoto2021statica} as B5 and B7 (Section~\ref{subsec:setup}). Each group is sorted by primary TTS$_{99}$.}
\label{tab:configs}
\setlength{\tabcolsep}{2.8pt}
\begin{tabular}{@{}llrrrrr@{}}
\toprule
Label & Rule & $S$ & $p$ & $t_r$ [ms] & Primary & Secondary \\
\midrule
O1 & Onsager-$\kappa T$ & 280 & 0.338 & 0.050 & \textbf{0.050} & 0.047 \\
O2 & Onsager-online & 360 & 0.326 & 0.059 & 0.053 & 0.057 \\
O3 & Onsager-$\kappa T$ & 320 & 0.428 & 0.057 & 0.057 & 0.039 \\
O4 & Onsager-online & 320 & 0.455 & 0.063 & 0.063 & 0.040 \\
O5 & Onsager-$\kappa T$ & 760 & 0.950 & 0.142 & 0.142 & \textbf{0.018} \\
O6 & Onsager-online & 960 & 0.933 & 0.173 & 0.173 & 0.025 \\
\midrule
B1 & TEC-style & 960 & 0.392 & 0.137 & 0.137 & 0.105 \\
B2 & plain SCA & 960 & 0.461 & 0.157 & 0.140 & 0.098 \\
B3 & plain SCA & 1560 & 0.439 & 0.199 & 0.199 & 0.132 \\
B4 & TEC-style & 1560 & 0.667 & 0.216 & 0.216 & 0.075 \\
B5 & plain SCA & 560 & 0.134 & 0.094 & 0.231 & 0.252 \\
B6 & plain SCA & 1560 & 0.819 & 0.271 & 0.271 & 0.061 \\
B7 & plain SCA & 1560 & 0.824 & 0.302 & 0.302 & 0.067 \\
\bottomrule
\end{tabular}
\vspace{-0.6em}%
\end{table}
\begin{table}[t]
\centering
\caption{Design evolution: all measured board implementations (v6: eight extractors; v7: multi-bit, with bias at $K{=}4$; Section~\ref{subsec:multibit}). TTS$_{99}$: best primary value per version; resources include the shell. $^{a}$/$^{b}$/$^{c}$Plus 792/1{,}560/1{,}566 URAM blocks.}
\label{tab:evo}
\setlength{\tabcolsep}{2.2pt}
\begin{tabular}{@{}lrrrrrrr@{}}
\toprule
Version & $E$ & MHz & cyc/step & cyc/flip & LUT & BRAM & TTS$_{99}$ [ms] \\
\midrule
v1 (HLS) & 1 & 300 & 40.2 & 0.52 & 0.20\,M & 141 & 0.634 \\
v2 (HLS) & 1 & 250 & 31.1 & 0.28 & 0.27\,M & 269 & 0.464 \\
v3 (HLS) & 3 & 225 & 31.1 & 0.28 & 1.05\,M & 890 & 0.211 \\
v4 (RTL) & 9 & 300 & 16.9 & 0.27 & 1.27\,M & 2{,}489 & 0.081 \\
v5 (RTL) & 12 & 275 & 16.9 & 0.27 & 1.68\,M & 3{,}318 & 0.069 \\
v6 (RTL) & 12 & 250 & 18.1 & 0.14 & 1.91\,M & 3{,}318$^{a}$ & \textbf{0.050} \\
v7 ($K{=}2$) & 12 & 250 & 18.1 & 0.14 & 2.30\,M & 3{,}318$^{b}$ & \textbf{0.050} \\
v7 ($K{=}4$) & 6 & 225 & 18.1 & 0.14 & 2.21\,M & 3{,}183$^{c}$ & 0.098 \\
\bottomrule
\end{tabular}
\vspace{-0.6em}%
\end{table}
\begin{table*}[t]
\centering
\caption{TTS$_{99}$ on the K2000 Max-Cut instance: a CPU baseline first, then hardware Ising machines in order of publication. Values of other machines are quoted from the cited publications; all CMOS and ReRAM results are simulations.}
\label{tab:comp}
\setlength{\tabcolsep}{1.5pt}
\begin{tabular*}{\textwidth}{@{\extracolsep{\fill}}l c c c c c c c c c c c c c c@{}}
\toprule
Machine & \multicolumn{2}{c}{Neal$^{a}$~\cite{dwave_neal}} &
CIM$^{b}$~\cite{doi:10.1126/science.aah4243} & SB$^{b}$~\cite{goto2019combinatorial} &
\multicolumn{2}{c}{STATICA$^{c}$~\cite{yamamoto2021statica}} &
bSB$^{d}$~\cite{goto2021high} &
\multicolumn{2}{c}{ReAIM$^{e}$~\cite{10609617}} &
GbSB$^{f}$~\cite{goto2026edge} &
DSSA$^{g}$~\cite{onizawa2026dssa} &
\multicolumn{3}{c}{\textcolor{anechoic}{\textbf{Anechoic}$^{h}$}} \\
\cmidrule(lr){2-3}\cmidrule(lr){4-4}\cmidrule(lr){5-5}\cmidrule(lr){6-7}\cmidrule(lr){8-8}\cmidrule(lr){9-10}\cmidrule(lr){11-11}\cmidrule(lr){12-12}\cmidrule(l){13-15}
Hardware & \multicolumn{2}{c}{CPU} & Optics & FPGA &
\multicolumn{2}{c}{CMOS} & FPGA & \multicolumn{2}{c}{ReRAM} & FPGA & CMOS & \textcolor{anechoic}{\textbf{FPGA}} & \textcolor{anechoic}{\textbf{GPU}} & \textcolor{anechoic}{\textbf{CMOS}} \\
Target cut & \multicolumn{2}{c}{33{,}000} & 33{,}000 & 33{,}000 &
\multicolumn{2}{c}{33{,}000} & 33{,}004 & \multicolumn{2}{c}{33{,}000} & 33{,}337 & n/s & \multicolumn{3}{c}{\textcolor{anechoic}{\textbf{33{,}000}}} \\
\midrule
$t_a$ [ms] & 4{,}610 & 5{,}646 & 5.0 & 0.5 & 0.13 & 0.48 & -- & 0.15 & 0.23 & 9.42 & 2.30 & \textcolor{anechoic}{\textbf{0.05}} & \textcolor{anechoic}{\textbf{0.11}} & \textcolor{anechoic}{\textbf{0.01}} \\
$P_a$ & 0.38 & 0.77 & 0.02 & 0.04 & 0.07 & 0.77 & -- & 0.47 & 0.8 & 0.989 & 0.98 & \textcolor{anechoic}{\textbf{1.00}} & \textcolor{anechoic}{\textbf{1.00}} & \textcolor{anechoic}{\textbf{1.00}} \\
TTS$_{99}$ [ms] & 44{,}413 & 17{,}693 & 1{,}139.74 & 56.41 & 8.23 & 1.50 & 0.26 & 1.11 & 0.68 & 9.61 & 2.7 & \textcolor{anechoic}{\textbf{0.05}} & \textcolor{anechoic}{\textbf{0.11}} & \textcolor{anechoic}{\textbf{0.01}} \\
\bottomrule
\end{tabular*}
\par\smallskip
\begin{minipage}{\textwidth}
$^{a}$SA on an Intel Core i9-12900KF, reported by ReAIM~\cite{10609617}.
$^{b}$Compiled by STATICA~\cite{yamamoto2021statica}; SB: Intel Arria~10 GX FPGA~\cite{goto2019combinatorial}.
$^{c}$Cycle-level simulation of a 2K-spin STATICA at 300\,MHz; the fabricated 65-nm chip has 512 spins~\cite{yamamoto2021statica}.
$^{d}$TTT at 99\% of the best-known cut ($\geq$\,33{,}004); $t_a$ and $P_a$ not reported~\cite{goto2021high}.
$^{e}$Simulation of a ReRAM processing-in-memory design, digital logic scaled from 130 to 32\,nm~\cite{10609617}.
$^{f}$Intel Agilex~7 FPGA at 591\,MHz; target: the best-known cut; $t_a$: 21{,}400 steps of 0.440\,\textmu s~\cite{goto2026edge}.
$^{g}$Post-layout simulation of a 28-nm processor at 500\,MHz, annealing phase only; target not stated (n/s)~\cite{onizawa2026dssa}.
$^{h}$Onsager-$\kappa T$, primary estimator (Section~\ref{subsec:tts}); $t_a$: mean round time; $P_a$: fraction of successful rounds. FPGA: AMD Alveo V80, 12 engines at 250\,MHz (O1, 171 rounds). GPU: NVIDIA GH200, 120 chains, own schedule (128 rounds; Section~\ref{subsec:gpu}). CMOS: simulation of 12 GlobalFoundries 22FDX engines, routed at 1.25\,GHz and cycle-exact to the FPGA in RTL (Section~\ref{subsec:asic}).
\end{minipage}
\vspace{-0.6em}%
\end{table*}

Table~\ref{tab:configs} lists representative configurations, all on v6: the best primary TTS$_{99}$ is \ttsP{} (O1, Onsager-$\kappa T$; at most 0.061\,ms with 95\% confidence), the best Onsager-online schedule reaches 0.053\,ms (O2), below its 17\% longer round time because 1 of its 171 rounds fails (Section~\ref{subsec:tts}), and the best secondary TTS$_{99}$ is \ttsS{} (O5).

\begin{figure}[t]
  \centering
  \includegraphics[width=\columnwidth]{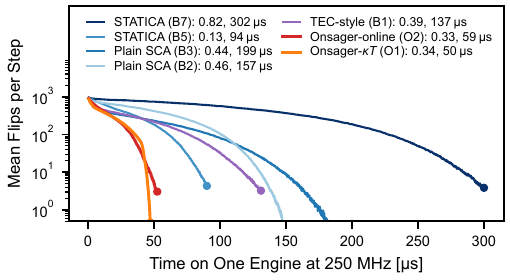}
  \caption{Mean flips per step over one trial for seven configurations of Table~\ref{tab:configs}, on a time axis from the cycle model (Equation~\eqref{eq:cycles}). The legend gives the on-board per-trial success probability and the mean 12-engine round time. The corrected forms succeed less often per trial than B1--B3 (0.33--0.34 against 0.39--0.46) but in rounds 25--43\% as long, and 2.5 times as often as STATICA's B5 in rounds 53--62\% as long.}
  \Description{Flips per step start near 900 for all seven configurations. The two Onsager curves end within about 50 microseconds, STATICA's short schedule at about 90, the plain SCA and TEC-style schedules at 130 to 180, and STATICA's long schedule at 300 microseconds.}
  \label{fig:flips}
\vspace{-0.6em}%
\end{figure}

Against STATICA's published schedules on the same board, the best corrected configurations are faster: O1 by $4.6\times$ on the primary estimator and O5 by $3.7\times$ on the secondary (against the better of B5 and B7). Against this work's plain-SCA and TEC-style schedules (B1--B4, B6), the factors are $2.8\times$ and $2.7\times$ (primary) and $3.4\times$ and $4.2\times$ (secondary).
Figure~\ref{fig:flips} shows where the gain comes from: (1)~O1 and O2 start cooler and finish in 280--360 steps instead of 560--1560; (2)~they flip fewer spins per trial (45--47\,k instead of 87--329\,k); and (3)~the engine's cost follows the flips (Equation~\eqref{eq:cycles}).

\subsection{Comparison with the state of the art}\label{subsec:sota}

Table~\ref{tab:comp} compares Anechoic with published results, at the 33{,}000 target except where the table lists another. On the primary estimator, Anechoic is \spdBSB{} faster than bSB, the fastest measured hardware~\cite{goto2021high}; with 95\% confidence, it is at least $4.2\times$ faster. That comparison mixes targets: bSB reports the same measure, as time to target (TTT), for a cut of at least 33{,}004, 99\% of the best known.

Figure~\ref{fig:alg} and Table~\ref{tab:comp} rank bSB differently because they measure different quantities. Figure~\ref{fig:alg} counts Monte Carlo steps: bSB needs the fewest, 293 steps to reach the target with 99\% probability, against 830 for Onsager-$\kappa T$ (Section~\ref{subsec:algcomp}). Table~\ref{tab:comp} measures time on hardware, where step costs differ: a bSB step multiplies the coupling matrix by real-valued amplitude changes, dense until the amplitudes settle, and GbSB takes 0.44\,\textmu s per step~\cite{goto2026edge}. A step of this engine processes only the 126--169 spins that flip and takes 0.15--0.18\,\textmu s (Section~\ref{subsec:update}). Part of the advantage is parallelism: the V80 runs twelve engines, and with O1 every 50\,\textmu s round has a success; one engine would need about 0.22\,ms (O5).

At the best-known cut, however, the proposed dynamics lag the state of the art: GbSB~\cite{goto2026edge} reaches 33{,}337 in 9.61\,ms. In a pre-registered study with the engine's floating-point model, the best Onsager-online schedule needs 54\,ms on twelve engines (95\% CI 44--65\,ms), 5.6 times GbSB's time, but its 16{,}000 steps exceed the 4{,}096-entry table, within which the best pre-registered schedule needs 105\,ms.

\begin{figure*}[t]
  \centering
  \includegraphics[width=\textwidth]{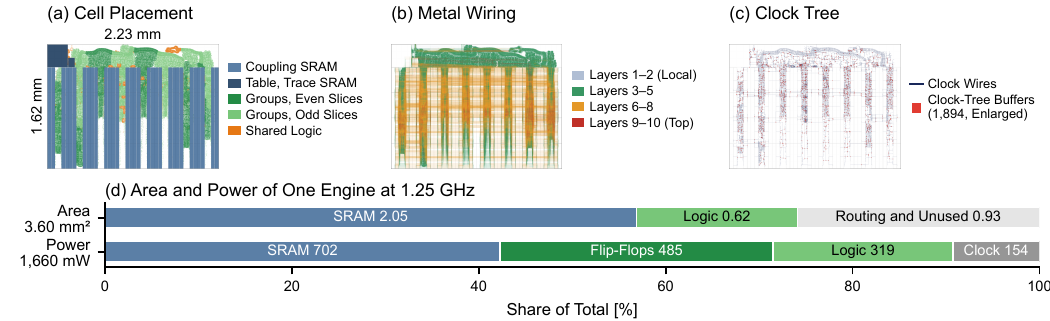}
  \caption{ASIC implementation of one engine (v6) in GlobalFoundries 22FDX, placed and routed at 1.25\,GHz. (a)~SRAM macros and standard cells by design hierarchy: each of the eight slices holds four groups next to their coupling macros, so every coupling read stays within its slice. (b)~Signal and clock wiring by metal layer. (c)~Clock tree: 2{,}798 buffers and 756 clock gates drive about 162{,}000 flip-flops (mean insertion delay 597\,ps, skew 81\,ps); enlarged markers show the 1{,}894 buffers that clock-tree synthesis placed. (d)~Area and power at the typical corner (0.80\,V, 25\,$^{\circ}$C); power is estimated from the signal toggle rates recorded in RTL simulation of one K2000 trial with schedule O1.}
  \Description{Four panels. (a) The die of 2.23 by 1.62 mm with tall blue SRAM columns and green group logic in eight vertical slices; dark blue table and trace memories at the top left and orange shared logic along the top. (b) The same die with dense green and orange wiring over the logic strips and orange horizontal wires over the SRAM. (c) Dark clock wires with red buffer markers over the logic strips and the top band. (d) Two stacked bars: the area is 57 percent SRAM, 17 percent logic and 26 percent routing and unused space; the power is 42 percent SRAM, 29 percent flip-flops, 19 percent logic and 9 percent clock tree.}
  \label{fig:asic}
\vspace{-0.6em}%
\end{figure*}

\subsection{Hardware cost and energy}\label{subsec:cost}
Board input power is sampled every 0.25\,s through the AVED Management Interface driver while all engines run back to back for 150\,s, over 99\% busy and without clock throttling. With O1, the board draws \pBoard{}, so the energy per solution, power times primary TTS$_{99}$, is \ets{} (Table~\ref{tab:platform}). The engines add 36--41\,W to the 66--73\,W the board draws at idle. The multi-bit images draw more on K2000: 126\,W for twelve 2-bit engines (6.4\,mJ per solution) and 94\,W for six 4-bit engines at 225\,MHz, which need 0.098\,ms (9.2\,mJ). Block RAM is the scarcest resource (89\%, Table~\ref{tab:hw}), because each engine stores its couplings four times (Section~\ref{subsec:layout}).

\subsection{The same engine on a GPU}\label{subsec:gpu}
To separate the platform from the algorithm, the engine is also implemented bit-exactly on an NVIDIA GH200~\cite{nvidia_gh200}. The GPU version reproduces the reference model, and therefore the V80, in 424 of 424 traced trials. Each chain, an independent anneal, runs on eight thread blocks that hold $J$ in registers and exchange flip masks via distributed shared memory.
Up to 240 chains run concurrently; the primary schedule is re-selected for the GPU (Table~\ref{tab:platform}).

\begin{table}[t]
\centering
\caption{The same bit-exact engine on three platforms (K2000, cut $\geq$ 33{,}000): the V80 and the GH200 GPU are measured, and the ASIC is estimated (Section~\ref{subsec:asic}); the GPU uses 120 chains for the primary and 240 for the secondary TTS$_{99}$. $^{a}$Board; engines: 39\,W, 2.0\,mJ per solution. $^{b}$Engines only.}
\label{tab:platform}
\setlength{\tabcolsep}{2.5pt}
\begin{tabular}{@{}lrrr@{}}
\toprule
& V80 & GH200 & ASIC \\
\midrule
Engines or concurrent chains & 12 & 120--240 & 12 \\
Step latency, one chain [\textmu s] & 0.15--0.18 & 0.63--0.72 & 0.031--0.035 \\
Primary TTS$_{99}$ [ms] & 0.050 & 0.114 & 0.010 \\
Secondary TTS$_{99}$ [ms] & 0.018 & 0.0078 & 0.0036 \\
Power [W] & 108$^{a}$ & 382 & 19.9$^{b}$ \\
Energy per solution [mJ] & 5.4$^{a}$ & 43 & 0.20$^{b}$ \\
Area [mm$^2$] & -- & -- & 43$^{b}$ \\
\bottomrule
\end{tabular}
\vspace{-0.6em}%
\end{table}

The V80 reaches the target \gpuSpd{} sooner on the primary estimator and uses \gpuEratio{} less energy per solution (GPU power from nvidia-smi). Its engines take 3.6--4.7$\times$ less time per step: about 250\,ns of every GPU step, more than a whole V80 step, is a message round trip between streaming multiprocessors. Twelve engines already succeed in every round, so more chains cannot shorten the primary TTS$_{99}$, but the GPU's 240 chains yield a $2.3\times$ lower secondary TTS$_{99}$. The same holds on the 51 G-set instances (Section~\ref{subsec:gset}) with schedules selected per platform: for Onsager-$\kappa T$, the V80 is 3.2 times faster on the primary estimator (on 47) and the GPU 1.9 times faster on the secondary (on 49), while the V80 (94\,W) needs 10 and 2.0 times less energy per solution, respectively (geometric means).

\subsection{ASIC estimate}\label{subsec:asic}

The eight-extractor engine (v6) is synthesized, placed and routed in GlobalFoundries 22FDX, a 22\,nm fully depleted silicon-on-insulator process, using 8-track low-threshold-voltage standard cells, with Cadence Genus and Innovus at the typical corner (Table~\ref{tab:asic}, Figure~\ref{fig:asic}). The memories are replaced by SRAM macros, and the logic is unchanged except that the step-table read takes two cycles, which changes no cycle count. As in the multi-bit engine (Section~\ref{subsec:multibit}), each coupling is stored once, in single-port SRAM (4 instead of 16\,Mbit per engine on the FPGA). In RTL simulation, this variant reproduces the FPGA engine bit- and cycle-exactly on eight test sets.

\begin{table}[t]
\centering
\caption{Routed resources and throughput (v6, 12 engines).}
\label{tab:hw}
\setlength{\tabcolsep}{3pt}
\begin{tabular}{@{}lrr@{}}
\toprule
& Per engine & Total (12 + shell) \\
\midrule
LUT & 149{,}512 & 1{,}906{,}208 (74.0\%) \\
Flip-flops & 131{,}555 & 1{,}767{,}564 (34.3\%) \\
BRAM (36\,Kb tiles) & 258 & 3{,}318 (88.7\%) \\
URAM & 66 & 792 (41.1\%) \\
DSP58 & 282 & 3{,}384 (31.2\%) \\
\midrule
Trials per second (O1, sustained) & 20{,}200 & 242{,}600 \\
\bottomrule
\end{tabular}
\vspace{-0.6em}%
\end{table}
\begin{table}[t]
\centering
\caption{ASIC estimate of one v6 engine in GlobalFoundries 22FDX (after routing, typical corner; power from the toggle rates of one K2000 trial with schedule O1 in RTL simulation).}
\label{tab:asic}
\setlength{\tabcolsep}{3pt}
\begin{tabular}{@{}lr@{}}
\toprule
Clock (setup / hold slack) & 1.25\,GHz (+7 / +191\,ps) \\
Die area & 3.60\,mm$^2$ ($2.23\times1.62$\,mm) \\
Cell area: SRAM / logic & 2.05 / 0.62\,mm$^2$ \\
Cells / flip-flops (clock-gated) & 866\,k / 162\,k (29\%) \\
Power (leakage) & 1{,}660\,mW (6\,mW) \\
Energy per trial (O1) & 16.4\,\textmu J \\
SRAM energy per applied flip & 0.15\,nJ \\
\bottomrule
\end{tabular}
\vspace{-0.6em}%
\end{table}
After routing, the engine meets 1.25\,GHz; smaller, faster SRAM macros keep the coupling read at one cycle. SRAM and flip-flops draw about 70\% of the power (Figure~\ref{fig:asic}d), because every applied flip reads a full 2{,}048-bit coupling row and only 29\% of the flip-flops are clock-gated (Table~\ref{tab:asic}). Because the variant is cycle-exact, twelve engines would reach the K2000 target 5.0 times sooner than the V80 (the clock ratio), in 10.1\,\textmu s, with 0.20\,mJ of engine energy per solution, a tenth of the V80's (Table~\ref{tab:platform}). This is 26 times below bSB's 0.26\,ms, the fastest measured hardware (Table~\ref{tab:comp}). On the ASIC, multi-bit couplings cost area, not engines or clock (Section~\ref{subsec:multibit}): with $K=2$ and bias, it routes at 1.25\,GHz in 5.8\,mm$^2$ (twelve: 0.35\,mJ per solution), and with $K=4$ it meets 1.25\,GHz in synthesis; both are bit-exact in simulation.

\subsection{Platform trade-offs}\label{subsec:tradeoffs}
The FPGA has the lowest measured primary TTS$_{99}$ and energy per solution (Table~\ref{tab:platform}), and one board runs every variant, up to $K=4$ with bias. However, its fabric clocks at 250\,MHz, a fifth of the ASIC's; on-chip memory limits it to twelve engines (six with $K=4$); and most of its 108\,W is idle power. The GPU needs no hardware design and has the highest measured throughput, but the round trip between multiprocessors in every step makes each solution slower and costlier in energy (Section~\ref{subsec:gpu}). The ASIC would be fastest and most efficient, but its coupling width is fixed at fabrication, and each extra bit costs area.

\subsection{Combinatorial optimization problems}\label{subsec:gset}
\begin{table}[t]
\centering
\caption{Measured speedup on 51 G-set Max-Cut instances~\cite{helmberg2000spectral,ye_gset}: how many times sooner the corrected rule reaches a cut within 1\% of the best known~\cite{benlic2013breakout,ma2017multiple} than plain SCA or TEC-style, from the twelve-engine TTS$_{99}$ on the V80 (geometric mean; in parentheses: instances on which it is faster). Extended grids of a pre-registered amendment.}
\label{tab:gset}
\setlength{\tabcolsep}{2.5pt}
\begin{tabular}{@{}lrrrr@{}}
\toprule
Speedup $\uparrow$ & \multicolumn{2}{c}{Onsager-$\kappa T$ vs} & \multicolumn{2}{c}{Onsager-online vs} \\
\cmidrule(lr){2-3}\cmidrule(l){4-5}
Graphs (instances) & plain SCA & TEC-style & plain SCA & TEC-style \\
\midrule
Random, $+1$ (15) & 1.68 (14) & 0.96 (6) & 1.32 (10) & 0.75 (8) \\
Random, $\pm1$ (10) & 2.21 (10) & 1.91 (10) & 1.64 (9) & 1.42 (7) \\
Toroidal, $\pm1$ (6) & 4.56 (6) & 3.83 (6) & 1.45 (6) & 1.22 (4) \\
Planar-like, $+1$ (12) & 2.71 (12) & 2.15 (12) & 0.82 (2) & 0.65 (1) \\
Planar-like, $\pm1$ (8) & 1.39 (7) & 1.52 (8) & 1.45 (8) & 1.59 (8) \\
\midrule
All 51 & \textbf{2.17 (49)} & \textbf{1.68 (42)} & 1.26 (35) & 0.98 (28) \\
\bottomrule
\end{tabular}
\vspace{-0.6em}%
\end{table}
\begin{table}[t]
\centering
\caption{Combinatorial optimization problems in software, with the baselines at their papers' settings (plain SCA and TEC: their K2000 settings, scaled per instance by a rule of this work). Steps: steps to reach a cut within 1\% of the best known with 99\% probability, geometric mean over the five instances every method solves (G6--G10); in parentheses: the number of the 20 solved, if fewer. Quality: closeness to the best known, from 0 (infeasible) to 1, after 4{,}000 (Max-Cut), 4{,}096 (GP) and 8{,}192 steps (TSP). $\uparrow$/$\downarrow$: higher/lower is better.}
\label{tab:cop}
\setlength{\tabcolsep}{2.5pt}
\begin{tabular}{@{}lrrrr@{}}
\toprule
& Steps $\downarrow$ & \multicolumn{3}{c}{Quality $\uparrow$} \\
\cmidrule(lr){2-2}\cmidrule(l){3-5}
Method & Max-Cut & Max-Cut & GP & TSP \\
\midrule
SA & 4{,}726 & 0.995 & 0.979 & 0.528 \\
aSB & 23{,}386 (7) & 0.148 & 0.000 & 0.000 \\
bSB & 879 (17) & 0.988 & 0.640 & 0.317 \\
dSB & 2{,}861 & 0.995 & 0.000 & 0.288 \\
ReAIM ASA & 43{,}187 (11) & 0.962 & 0.567 & 0.416 \\
APC-SCA & 5{,}875 (15) & 0.960 & 0.166 & 0.542 \\
plain SCA & 4{,}594 (11) & 0.542 & 0.000 & 0.000 \\
TEC & 92{,}387 (19) & 0.986 & 0.433 & 0.052 \\
Onsager-online & 1{,}729 & 0.990 & 0.026 & 0.474 \\
Onsager-$\kappa T$ & 2{,}407 & 0.994 & 0.026 & 0.476 \\
\bottomrule
\end{tabular}
\vspace{-0.6em}%
\end{table}
Table~\ref{tab:gset} extends the hardware evaluation to sparse graphs: the 2-bit engine (Section~\ref{subsec:multibit}) measures all 51 G-set instances~\cite{helmberg2000spectral,ye_gset} that fit it ($N\leq2048$). Because the papers of plain SCA and TEC give no G-set settings, every rule, baselines included, takes its schedules from the same pre-registered selection on the engine's floating-point model (32-point grids, 64 tuning and 256 held-out runs per step budget), with all parameters scaled to each instance. Onsager-$\kappa T$, the form of the fastest K2000 schedule, is faster than plain SCA on 48 of 51 instances ($1.99\times$ on geometric mean) and than TEC-style on 41 ($1.55\times$) with the pre-registered grids. With the extended grids of a pre-registered amendment (Table~\ref{tab:gset}), it is faster on 49 and 42 ($2.17\times$ and $1.68\times$), with a geometric-mean TTS$_{99}$ of 0.11\,ms against 0.24 and 0.19\,ms; its largest gains are on the toroidal graphs. The model predicted 50 and 40 wins, and the measured success rates agree with it within 1.96 standard errors plus 0.02 in all 264 instance--schedule cohorts. Onsager-online gains less ($1.26\times$ and 0.98--$0.99\times$) and is slower than both baselines on the planar-like $+1$ graphs.

Table~\ref{tab:cop} widens the comparison in software to the three problems of ReAIM~\cite{10609617}: Max-Cut on G1--G20, GP on G1--G5 and G14--G17 (scored against the best partitions found by SA with balance-preserving swaps) and the TSP on six TSPLIB instances~\cite{reinelt1991tsplib}, in Lucas's encodings~\cite{lucas2014ising} with fixed penalties and the linear terms as biases. The couplings are unquantized; matching full precision would take 6-bit couplings for the TSP, against 2 bits for Max-Cut. As on K2000, only the two corrected rules are tuned. \emph{Max-Cut:} on the five instances that every method solves, only bSB needs fewer steps than the corrected rules, but over all 20, SA and dSB need 0.44--0.62 times as many, so on sparse graphs any advantage of the engine over them must come from its cheaper steps; bSB misses the target on 3 of the 20 instances, and plain SCA and TEC, at their K2000 settings, on 9 and 1. \emph{GP and the TSP:} neither plain SCA nor a corrected form reaches the target, in software or on the 4-bit GPU kernel, and the V80 reproduces these runs bit for bit (Section~\ref{subsec:multibit}). The balance penalty of GP couples all spins, so parallel updates flip about half of them per step: only 4--5\% of the corrected rules' runs end balanced, and none of plain SCA, against all runs of SA and 58--66\% for ReAIM ASA and bSB. On the TSP, the corrected rules return valid tours, where plain SCA returns none, of quality 0.47--0.48 (random tours: 0.37; best baselines: SA 0.53, APC-SCA 0.54); only ReAIM ASA and bSB reach the target, once each.

\subsection{Limitations}\label{subsec:limits}
Synchronous rules fail on constrained problems such as GP and the TSP, whose penalties couple all spins, so that many flip together and overshoot (Section~\ref{subsec:gset}). The ASIC results, like the CMOS and ReRAM designs in Table~\ref{tab:comp}, are simulations, not silicon measurements (Section~\ref{subsec:asic}).

\section{Conclusion}
In synchronous annealing, the previous state of each spin returns to it through its neighbors, an echo that earlier parallel annealers suppress with tuned pinning or time-step couplings. Anechoic subtracts the echo: dynamical mean-field theory gives its coefficient, which the engine counts with one population count per step (Onsager-online) or approximates from the temperature (Onsager-$\kappa T$). At every step budget on K2000, both forms reach lower Ising energy than five published binary-spin algorithms run at their papers' settings. On the AMD V80, twelve flip-proportional engines (\cycStep{} cycles per step plus \cycFlip{} per flip) turn the saved steps and flips into time and reach the K2000 target in \ttsPr{}: \spdBSB{} faster than the fastest published hardware~\cite{goto2021high} and $4.6\times$ faster than STATICA's schedules on the same board. With 2-bit couplings, Onsager-$\kappa T$ is also faster than plain SCA on 48 of 51 G-set instances. The same bit-exact engine needs \gpuSpd{} more time and \gpuEratio{} more energy per solution on an NVIDIA GH200, though the GPU's 240 anneals give 2.3$\times$ higher throughput. As a GlobalFoundries 22\,nm ASIC at 1.25\,GHz, twelve engines would reach the target in about 10\,\textmu s, $26\times$ faster than bSB, with 0.20\,mJ of engine energy per solution.

\section*{Use of AI Tools}\label{subsec:ai}
Artificial intelligence (AI) tools such as Gemini~3.1~Pro, Claude Opus~5.5 and GPT-6 Astra are used to proofread the manuscript and to check the related work and the theory, including the machine-checked proofs of Section~\ref{subsec:echo}. Additionally, they are used to re-implement the published algorithms used for comparison (Section~\ref{subsec:algcomp}), while all values reported by other papers, including those in Table~\ref{tab:comp}, are quoted directly from the cited publications.

\bibliographystyle{ACM-Reference-Format}
\bibliography{refs}

\end{document}